# Institution:

Towson University - Graduate School

# Class:

Applied Information Technology – AIT_710-102

# Case Studies:

Information Security and Assurance

# Research Topic:

Cloud Security Architecture and Implementation
A practical approach

# Submitted By:

Maxwell Farnga, CSAA, OCS, MCP

**Certified and Experienced Professional in:**

Cloud, Databases, Security, IT Operations, DevOps, IaC, Infrastructure Engineering and Architecture

# Submitted to:

Prof. Friedman

# Advisor

# Submission Date:

May 14, 2018

# Cloud Security Architecture and Implementation

## Contents

## Abstract
While cloud computing provides lower Infrastructure cost, higher agility and faster delivery, it also presents higher operational and security risks for business critical assets; but a well-designed solution and security architecture will keep businesses safe during and after migrating their assets to the cloud.

Cloud computing has become the most popular technology trend that many small, mid-size and large organizations are enticed to invest into or at least evaluate. It is the process of storing and accessing the company's data, software and other technologies from a networked location, which is managed by another company called a cloud provider. While moving to the cloud appears to be more cost effective than investing in additional hardware and software to increase efficiency, it comes with some additional security considerations, because the consumer may be sharing the technology platforms with competitors and the public.

This paper will research and identify best security practices and how to improve a security architecture in an enterprise cloud environment. It will also review some of the cloud vendor's practice statements, other cloud research literatures, cloud reference architecture and other cloud security frameworks.

Cloud computing was found to have some high security risk and higher operations cost associated with it. Therefore, this research will provide some recommendations on best security practices and architecture to allow businesses make more informed decisions and implement more secure and affordable designs for their cloud environments.

## Cloud computing overview and evolution
The National Institution for Standards and Technology (NIST) defined cloud computing as "a model for enabling convenient, on-demand network access to a shared pool of configurable computing resources (e.g., networks, servers, storage, applications, and services) that can be rapidly provisioned and released with minimal management effort or service provider interaction" [4].
In recent years, cloud computing has emerged as a buzzword in many corporations (small to large) and in the Information and Communication Technology industry as a whole. It gives an organization the option to extend their traditional data Center or abandon those data centers for a more advanced Data Center that comes with higher computing power, high agility, high efficiency, speed to market, lower administrative overhead, and measured operations cost. The concept of cloud computing has been around since the late 1960s and early 1970s, but it started to gain more momentum in 1999 when saleforce.com released their cloud services, followed by the development of Amazon Web Services in 2002.

To help the readers better understand the concept of cloud computing, let's take a step back and present a few popular analogies. Professionals often compare cloud computing to an electric grid and other utilities. Dating back to the early 18 century, many individuals and businesses had their own power sources to produce electricity; smaller businesses had smaller generators based on their needs and larger corporations had extremely large generators; some large manufacturing plants even needed a mini dam to satisfy their electricity needs. Imagine the cost, administrative overhead and how time consuming it was to manage the generators and other power sources which includes fueling, maintenance, extra staff, equipment cost and other operations cost; meanwhile electricity supply is not their core business. [7]

Fortunately, the first few commercial electric power grids evolved between 1879 through 1915, and ever since life has been better for residential and industrial electricity consumption. [10] Because of this development there are no isolated electric power sources in every home or corporate building. Now they just need an electrical breaker and a thermostat to control the electricity, and only pay for the amount of electricity utilized. The same examples can be used with gas, water supply, telephone, and other utilities. [7]

This concept can also be applied to cloud computing technologies, as it is fully virtualized and hides the backend and physical operations from the consumers. In traditional IT operations, organizations spend millions of dollars building and managing large data centers to ensure that their applications and data are secure, proficient and always available for their business users and customers. The emergence of cloud computing (computing as a utility) does not only provide flexibility and efficiency for organizations, it also helps them measure operation costs and leverage tools that are not easily accessible in a traditional data center.

### The evolution of cloud computing

Prior to cloud computing, organizations spent days, weeks and sometimes months to provision new servers for projects, new systems, software upgrades, performance and scalability reasons. This process was extreme long because IT departments needed to order new hardware, and the hardware vendors would take days to prepare and ship them, the shipment will take a few days to arrive, then the IT team need to register, image and configure the new servers, and finally it will be handed over the appropriate team that requested them. This would take at least a week or two depending on the urgency; in larger organizations with more bureaucracy this might take 2-6 months to fulfill similar requests.

Because of the above delays and limited flexibility, the Technology Industry and businesses have been craving for ways to utilize high power and on-demand computing along. Because of these needs, computing has evolved from a single operating system per hardware, to virtualization (single hardware multiple operating systems) and now it has advanced to cloud computing (on-demand computing as services and no hardware required). [7]

### Types of cloud services and architecture models

Since cloud technologies are addressing a wide range of demands, delivering a one-size-fit-all solution would not have worked, because every organization is different from the other in multiple ways. In this section, we will discuss the different layers of cloud services as well as their deployment models.

Because cloud computing is more than just hosting an organization's information system at a network location, it encompasses prebuilt services not otherwise available in a traditional data center. Therefore, it has evolved to handle the diverse needs of the different consumers and businesses who need the service. The cloud services can be divided into two major categories, namely the Deployment Model and Service Model. [7]

### Deployment Model includes:

#### Private Cloud

In this deployment model, the cloud environment is hosted exclusively for a particular organization either on premise or off premise, and it can be managed by the organization or a third party [4].

An example of this deployment model could include: an organization installing OpenStack, CloudStack or other cloud like management services to convert an existing data center into a cloud environment. This can be used by large organizations that already have a well-established IT Department with a lot of critical applications and data.

This model is a lot more secure, because the organization can use their traditional security policies and network architecture to properly isolate, manage, and control their IT Infrastructure.

### Hybrid Cloud
This is a composition of two or more cloud environments that remain separate entities but are connected through proprietary technologies to enable them to seamlessly share data and applications through a standardized process. [4]

An example of this could be an organization using a public cloud like Amazon Web Services and connecting it to their current data center using technologies like Amazon Direct Connect to extend their Services like Active Directory and other applications.

### Community Cloud
In this deployment model, the cloud infrastructure or platform is shared by a specific group of organizations with common objectives or mutual concerns. This isolated and shared system is on a dedicated hardware and can be hosted by one of the organizations, or a third party provider. [4]

An example of a community cloud include; the US intelligence community which includes multiple agencies like NSA, CIA, FBI, and DoD community cloud hosted at one of the agencies location and it is managed by a major cloud provide.

### Public Cloud
In this deployment model, the cloud services are made available to the general public, businesses and large institutions, and it is owned and operated by a Cloud provider who sells the services as a utility for consumption. [4]

This is the most popular cloud deployment model and many organizations and average users are aware of the cloud services offered in this model. In the below section, we will expand on the details of this model. There are thousands of public cloud providers and services, and examples include: Amazon Web Services, Microsoft Cloud services, Google, Salesforce, DropBox, Apple Cloud, Samsung cloud and many more.

### Services Model includes:
Because of the rapid advancements in technologies and increasing demands of organizations to rapidly provision business systems and applications in limited time frame; it is extremely slow, costly and less secure for organization to always develop custom applications, acquire new hardware to deploy new systems or assemble teams to recreate wheels that are already in working for others. This is where the shared economy comes in with public cloud computing. Because of the high demands and the diverse needs, public cloud computing is offered in a tiered service model to satisfy these demands.

Although the cloud systems are managed by the cloud provider and primarily consumed by the cloud customers, there is a shared operational, security and compliance responsibilities between the providers and the consumers. This is also known as "Security **of** the Cloud" for the Provider's responsibilities and "Security **in** the Cloud" for the Consumer's responsibilities.

### Software as a Service (SaaS) - Benefits and security considerations
In the SaaS service model, the cloud providers deliver applications directly to end users and businesses for consumption. The consumers do not have visibility to the underlined infrastructure and the maintenance process, because everything is abstracted. They only have access through a web browser or mobile application, and can only change their specific settings within their applications.

Examples of SaaS can include Email services, Dropbox, Apple ID, ADP payroll services, and Jira. These services are typically used by individuals and businesses that do not have the time, resources or expertise to deploy a customer or more advanced solutions on premise.

With regards to the shared responsibilities, the users are only responsible for managing the users, credentials and settings within the application; while the providers manage the buildings, Network, Storage, Server Hardware, Hypervisor, Operating System, Database, Middleware server, Security, Integration, Runtime environment, and Application layers. In this service tier, if the application is exploited or compromised because the consumer did not properly manage their credentials, or did not disable user account for a separated employee, the cloud provider is not liable for any resulting incidents.

### Platform as a Service (PaaS) - Benefits and security considerations
In the PaaS service model, the public cloud providers deliver the underline infrastructure, middleware platform and runtime environments for the consumers to deploy and run their applications. The consumer may be aware of the kind of database, middleware server or runtime environment, to properly configure their application software to operate, but they do not manage or maintain those systems.

Examples of PaaS can include AWS Lambda, Elastic Beanstalk, Google App Engine, Apache Stratos, and RDS. These platforms were typically used by organizations or developers who want to run their custom or COTS applications without out worrying about the operating environment.

In reference to the Shared responsibility between the consumers and providers, the consumer needs to ensure that their code is functional and compatible with the platforms they are deploying on; they also need to ensure that their applications do not have any security vulnerabilities and they need to properly manage access to these application. In this model the consumer need to also understand their application architecture and codes, otherwise the cloud provider may not support any application level issues or vulnerabilities that erupt.

### Infrastructure as a Service (IaaS) Benefits and security considerations
The IaaS is the most complex and the most vulnerable cloud service delivery model, and it will be the most discussed cloud service tier in this research paper. This service tier allows the consumers (small to large organizations) to provision Network, Virtual Machines, Storage technologies, Load balancer, and other infrastructure services along with the platforms, applications and integration needed to enable their business needs and functions. Unlike the other two services tiers (SaaS and PaaS), the consumers need more technical skills and abilities, and they are aware of most of the technologies that enable their solution; except they do not have physical access to the facilities, hardware and other technologies that are being used.

Examples of the IaaS include AWS VPC, EC2, Route 53, Google Compute engine, Microsoft Azure Compute cloud, and many more. These Infrastructure services are utilize by small organizations starting up with/without the initial capital to invest in an on premise data center environment  or a midsize organization with a single data center and looking to use public cloud as their Disaster recovery site rather than investing capital into a building another data center, or up to a large enterprise with multiple data centers but looking to reduce their physical footprint and utilize the public cloud services to increase agility and high power on demand computing.

In this service model, the consumers have lot more responsibilities including: cloud architecture designs, security, operations, performance, high availability, resilience, data management, and network traffic. Due to the technical complexities and high security vulnerabilities, the consumers need in house experts who understand the providers' cloud infrastructure and various services to properly design the right solution that will meet the customer's responsibilities listed above.
The IaaS security vulnerabilities, threats and counter measures will be further discussed in the risk management section of this paper.

### Cloud computing practice statements overview
Given the fact that the cloud computing services are opened to the general public, there is a high chance that there will be some bad actors, among the consumers with good intentions. So cloud providers have implemented very strong restrictions, governance and industry best practices to help keep every consumer safe on their various cloud platforms.

These policies and standards are in place to help govern the consumer's use and protect the cloud provider's services, as part of the providers' shared security and operational responsibilities in the cloud. Below are some term of services and End User Agreements from two of the largest cloud services providers, who have services that span across the three service delivery model discussed earlier.

The below statistics represent the distribution cloud consumers across the major service tiers and the major cloud providers:

According to over 300,000 responding organizations who use cloud computing; 61% use SaaS, while PaaS users represent 39%, IaaS users represent 53% and other cloud services represent about 28%. [3]

Among the thousands of cloud providers, below are some of the major providers who services span across the various cloud services tiers; based on over 300,000 respondents; 45% use Amazon Web Services, while 39% use Microsoft Azure, 18% use VMware, 11% use Rackspace, 8% use IBM and others 5% or less. [3]

### Cloud security Risk Management
We have discussed the various flavors of cloud computing, the shared security and operational responsibilities between the consumers and the vendors, as well as some of the vendors' approach to managing security and governance for their services in the cloud.

Regardless of which deployment model (Private, Community, Hybrid or Public) the consumer choose, there will be lot of security implications and risk to manage; even traditional IT environments have security challenges. In this paper, our focus will be on Public cloud deployment models, because it the most vulnerable and have the most responsibilities for the consumers, under the shared security model.

Irrespective of the Public cloud service tier (SaaS, PaaS or IaaS), the consumers, especially enterprise organizations, assume a level of operational and security risk; the more responsibilities the higher the risk. So the consumers need to have the right skills in house to design a well architected cloud environment that meets the following pillars: Security, Resilience, Performance, and appropriate standards.

### Cloud computing threats and vulnerabilities

In information security, **vulnerability** is a weakness that could exist at the network, hardware, application, policy, people and other systems; and these weaknesses could be exploited by an attacker to cause harm or for personal gain. While a **threat** is any actor, with malicious intent or unintentional events that could exploit potential vulnerabilities and can have devastating impact including loss of services and/or financial impact.

To ensure that organization cloud environments and data are adequately secured, there are several vulnerability management tools, controls and strategies that can help organizations assess their cloud environment on an on-going basis. Below is a summary of some possible vulnerabilities and threats that can compromise the Confidentiality, Integrity and Availability of a cloud computing environment.
- Default credentials and/or weak passwords
- Database server running with lower security setup
- Operating Systems, middleware and/or applications running with outdated patches or unused ports
- Web application database with SQL injection vulnerabilities
- And other CVEs based on a known vulnerability database

To properly manage the risks in a cloud environment a comprehensive cloud governance framework is a must; and developing this framework requires a thorough understanding of the various threats and vulnerabilities landscape and the appropriate controls, processes and tools necessary to mitigate the risks. [2]

Below is a Risk assessment chat that will be used to categories the potential Risk in a cloud environment.

| Table 2: Risk Assessment for cloud computing environment | | | | |
|---|---|---|---|---|
| | Risk Impact – Severity Category | | | |
| Probability of Vulnerability | **Negligible (A)** | **Marginal (B)** | **Critical (C.)** | **Fatal (D)** |
| **Improbable (1)** | Low -1A | Low – 1B | Medium – 1C | Medium – 1D |
| **Probable (2)** | Low – 2A | Medium – 2B | Medium – 2C | High – 2D |
| **Occasional (3)** | Medium - 3A | Medium – 3B | High – 3C | High – 3D |
| **Frequent (4)** | Medium – 4A | High – 4B | High – 4C | High – 4D |

### Vulnerability:

Below are some of the many vulnerabilities associated with migrating a company's assets to a cloud environment: This section will also recommend some counter measures based on NIST special publication 800-53r4, which provides standards for the Federal Information Systems Management Act.

**Session Riding and Hijacking (Risk level = 4D)**
In session hijacking, the attackers use a valid session key to gain unauthorized access to an application and other information systems running on the cloud. This can be done by stealing the user's cookie from one session and using it on another session; if the attacker succeeds in riding on the remote session they can maliciously cause a lot of damage, including: illegal transactions, deleting user's data, disclosing private data, sending spam communications, or even executing commands to change the systems settings and making the system more vulnerable or impacting other user's account. [16]

*Recommended risk mitigation:*
AC-10 (CONCURRENT SESSION CONTROL) – Mitigates the risk by limiting the number of concurrent sessions for privileged and non-privileged users system wide. [17]
AC-11 (SESSION LOCK) – Implement session lock to terminate user's session when inactivity is detected for a defined period. [17]
SC-23 (SESSION AUTHENTICITY) – Protects information communication at the session level and establishes confidence between both ends of the communication, by continually identifying and validating other parties. This helps against man-in-the-middle, session hijacking and other cross-site scripting attacks. [17]

HIPAA 4.14. Access Control - also recommends (Automatic Logoff and Encryption and Decryption) of sessions whenever Electronically Protected Health Information (ePHI) is involved. This will automatically terminate a session after a predefined period of inactivity. [5]

In addition to AC-10, AC-11 and SC-23, consumers need to utilize multi-factor authentication (MFA) in their cloud applications. MFA should be a requirement for critical applications and for privileged users.

**Virtual Machine Escape (Risk level = 2D)**
Hosting multiple virtual machines on the same hardware can present significant vulnerability that could allow mischievous users, who are sharing the same hardware with others, to run malicious codes that will exploit the data in process at the hypervisor level. Data at rest and data in flight can be secured with encryption, but data in processor cannot be protected with encryptions. Because the exploit is based on an Intel vulnerability, and it can allow the attacker to access data being processed by the central processing unit (CPU). When a VM escapes, the attackers can remotely access and/or read other VM data as they are being processed through the CPU. [16]

*Recommended risk mitigation:*
SI-7 (1)(2)(3)(15)(16) - (SOFTWARE, FIRMWARE, AND INFORMATION INTEGRITY) – Recommends implementing a number controls to ensure the integrity of processes running on an information system. These controls can include verifying the firmware is up to date, that it is digitally signed, contains no vulnerabilities, and notifying if anomalies exist and an automated response to any incidents. [17]
SI-14 - (NON-PERSISTENCE) – This control mitigates the risk from advanced persistent threats by limiting the window of opportunity for cyber-attacks, and it provides a known state for computing resource for a limited execution time for selected systems, so attacker is not able to exploit any possible vulnerabilities that may exist. [17]

In addition to SI-7, SI-14 and encrypting every sensitive hard drive, consumers and providers need to ensure that there is a threat intelligent and host base intrusion detection and prevention system running on the hardware. Most importantly, the need to effectively monitor the CVEs and apply required patches immediately.

**Reliability and Availability of Service (Risk level = 2C)**
Reliability, high availability and resilience are key components of the CIA in cloud security. Since the consumers do not manage the physical infrastructure, relaying on a single cloud provider or service could cause their business to encounter unexpected outage when the cloud provider's services experience any downtime. [16] Some classic examples of relaying on a single service include: the outage of Amazon S3 in early March 2017 and September 2017, when S3 brought down many popular websites and other business services. Atlassian (a large SaaS provider) hosts most of their services on AWS, especially Bitbucket. Also in March 2018, many companies experienced outage with their code repositories on Bitbucket, and this outage was due to an issue with AWS Direct Connect service in Northern Virginia.

*Recommended risk mitigation:*
CP-11 (ALTERNATE COMMUNICATIONS PROTOCOLS) - Recommends an alternative system to help with business continuity in case the primary systems fail. [17]
SI-13 (PREDICTABLE FAILURE PREVENTION) - Determines the mean time to failure (MTTF) for specific environment of operation; and provides a substitute system components and a means to exchange active and standby components that meets the (MTTF). [17]

Based on the importance and sensitivity of the system to the business operations, the consumers need to design or ensure that the cloud provider has a high availability infrastructure in place. For example, if the consumer hosts majority of their systems in the cloud and they need high availability and high resilience, provision the resources in multiple availability zones. If it is a global service like S3, have a standby system on Azure cloud services.

**Insecure Cryptography (Risk level = 3C)**
In a cloud environment, the consumers need to identify and classify the data that needs to be protected at rest, next determine if the threat is insider or external users; then encrypt the data or the entire drive wherever the data resides. However, many cryptographic algorithms can be maliciously decoded by attackers, it might take them longer, but crucial flaws make these algorithm implementation vulnerable to threats. "Although utilization of those VMs into cloud providers' data centers provides more flexible and efficient setup than traditional servers, but they do not have enough access to generate random numbers needed to properly encrypt data". [16]
The biggest problem with insure cryptographic often includes not encrypting data that need to be encrypted, insecurely storing the key, generating weak keys or not rotating existing keys.

*Recommended risk mitigation:*
SC-13 (CRYPTOGRAPHIC PROTECTION) – employs cryptographic protect and control access to classified, non-classified and other sensitive information. This control limits information and resources access to only authorized users. [17]
SC-28 (PROTECTION OF INFORMATION AT REST) – Addresses the protection of confidentiality and integrity for information at rest, which includes users, systems and operational information. This layer of protection includes: the use of cryptographic mechanisms, files share scanning and Write-Once-Read-Many WORM tools. [17]

The consumers should audit the SaaS provider or have them confirm that they are encrypting sensitive data according to SC-13 and SC-28. In addition, consumers using the PaaS or IaaS levels, should encrypt all hard drives where possible or at least encrypt all drives that hold business critical/sensitive information like PII, Credentials, Databases data, application configurations and others. The second aspect is to use table level encryption for all credit card numbers, social security, electronic protected health information (ePHI), and other very sensitive data. Lastly, remember to properly secure encryption keys and do not use it for everyday logins to the servers. Connect all servers to active directory or create local logins with remote logon privilege. Though encryption algorithms can be compromised by attackers, having encryption is better than having no encryption, because it will delay the attack. Also remember to generate stronger encryption algorithms and rotate them frequently, based on the organization's retention period.

**Vendor Lock-in, Data Protection and Portability (Risk level = 2C)**
This vulnerability occurs when the consumer is not able to transfer their applications and services from one provider to another, due to the heterogeneous standards and services offered by providers.
Data protection and portability is also vital, because organizations need to have the ability to move their data and critical systems from one provider to another or even back on premise without reinventing the wheel or significant development efforts; this is another weakness that still needs to be addressed in cloud computing. In a situation where the contract between the consumer and provider is terminated for whatever reason. For example, consumer no longer interested or trust provider, or provider goes out of business, or any unexpected situation. The consumer need to be assured that they can move their data from one provider to another, and that whatever data that is left of the provider's system is not misused, but safely deleted after migrating the original copies. [16]

*Recommended risk mitigation:*
SC-27 (OPERATING SYSTEM-INDEPENDENT APPLICATIONS) - Recommends designing applications to be Platform agnostic, so the application can be portable from one system to another. [17]

The same applies to cloud systems. Many providers have multiple services to manage, automate, secure, store, scale and secure the consumers' applications and data in the cloud, but many of these services cannot be used at another provider, especially at the IaaS level. To address this concern, consumers need to be smart and utilize services that are independent and work across multiple cloud providers. For example, instead of using AWS CloudFormation to manage Infrastructure automation, use alternative tools like Ansible, Puppet, or Chef, because they can be used on premise and across many cloud platforms.

**Internet Dependency (Risk level = 3A)**
Since cloud computing is fully dependent on the internet, consumers need high performance internet to access their data and systems in the cloud environment. Without reliable internet connection, the consumers cannot effectively benefit from the services and functionalities offered by the cloud computing.

*Recommended risk mitigation:*
In cases like these, the consumers need to make a rational decision to move their data to the cloud or not. E.g. A small clinic or hospital in a rural area with limited internet connectivity may choose to continue hosting their local clinical applications rather than moving it to a cloud environment.

HIPAA 4.7 Contingency Plan, recommends the establishment of a contingency plans to help the organization recover their systems and continue operation, in case of any incidents or major interruptions. [5]

In such case, the rural organization could use the cloud as a backup and recovery point, that way the business operations are not impact by the internet connectivity and they also save the cost from building another infrastructure for disaster recovery.

### Threat:
### Abuse and Nefarious Use of Cloud services (Risk level = 4A)
Because cloud services are opened to the public, anybody can use them with little or no verification process during registration, and some cloud providers have weak or no fraud detection mechanisms in place. So it is easier for any consumer including malicious users like spammers, malware authors and other cyber-criminals to take advantage of these vulnerabilities and sign up for cloud accounts, to execute their clandestine commands. For example, if a malicious IaaS user executes a malware in their cloud instance, and it affects the host, it could also be transmitted to other cloud instances through machine-to-machine traffic. [16]

*Recommended risk mitigation:*
AC-6 (LEAST PRIVILEGE) – Mitigate misuse of both insider and external users by employing the principal of least privilege for all authorized users. This gives users the minimum level of access required to perform their basic functions in the systems. [17]

SI-3 (MALICIOUS CODE PROTECTION) – Employs malicious code detection and protection at all entry and exit points to eradicate malware codes from the firewall, servers, proxy, file-storage and other critical information systems. [17]

SI-4 (INFORMATION SYSTEM MONITORING) – Continuously monitor the information system for any attack indicators, malicious codes and other vulnerabilities, and immediately alert administrators and automatically respond, if any are detected. [17]

Consumers should confirm that the SaaS provider has implemented the AC-6, SI-3, SC-4 and other security procedures to reduce or eradicate misuse of their services. This way, every new user who signs up for services can only do the minimum level of tasks, such that they would not be able to perform any action beyond their account. Also verify that the provider has a threat intelligence system in place (preferably AI capable) to quickly detect fraud and automatically respond.

### Insecure Interfaces and APIs (Risk level = 3C)
With cloud computing, the need for interacting with other systems to share, integrate, and make the consumer data more portable from one system to another has increased. To enable this integration capability, it requires the use of Application Program Interfaces (APIs). These APIs can also present new security challenges if they are not properly secured. "These APIs need to be secured because they play an integral part during provisioning, management, orchestration and monitoring of the processes running in a cloud environment." Common examples include: developing APIs and plain text credentials in the code or transmitting data without SSL. [16]

*Recommended risk mitigation:*
SA-15 (DEVELOPMENT PROCESS, STANDARDS, AND TOOLS) - recommends that organizations use best practice development processes, standards and tools to address security requirements, standardize development process, and manage configurations.
This will help eliminate APIs that are vulnerable to cyber-attacks. [17]
Also use data encapsulation, variables, objects, encrypted credentials in the API codes and use SSL to encrypt data in flight.

### Insider threat (Risk level = 3D)
In a traditional IT data center environment, the consumer would have full control over the physical, logical and data security (For, example, building, server hardware, hypervisor, network and others), therefore, they can design and implement their own defense in-depth strategies and policies to mitigate any form of insider threat. But considering the fact that cloud involves outsourcing critical business data and applications to the providers' premise; which means the consumers lose full control over the physical security and operational aspect of their underlying infrastructure. These insider threats could involve a privileged employee intentionally or inadvertently causing harm to the system(s) that could result in an unauthorized disclosure, data integrity, systems outage, or other financial lost. [16]

*Recommended risk mitigation:*
AT-2 SECURITY AWARENESS – recommends regular security awareness for employees, so they can be aware of basic security risks and procedures. [17]
PM-12 INSIDER THREAT PROGRAM - leverages a team from across major departments to help manage, prevent, detect and respond to any form of insider threats. [17]
AC-5 SEPARATION OF DUTIES – recommends the separation employees' responsibilities, to ensure proper checks and balances. [17]

In addition to having AC-5, AT-2, PM-12 and other control mechanisms, the consumers must also have an appropriate management of authentication, authorization and auditing.

In addition to the above controls, PaaS and IaaS consumers should encrypt their hard drives to avoid unauthorized disclosure if the consumer's data is accessed directly from the hardware. They should also setup redundancies across multiple availability zones, to failover when a resource is intentionally or accidentally interrupted by an insider.

### Data Loss and Leakage (Risk level = 2D)
Data loss or leakage could result from operational and security related threats. A typical operational data loss would include disk failure, server corruption or an accidental deletion of a resource that holds your data. In situations like these, the consumers would lose data if there is no recovery process in place or if there is no measure to mitigate failure of their systems.
Data leakage is a security related threat, and it occurs when the consumers do not properly manage their encryption keys or credentials for their environment. If these keys or credentials are discovered by the attacker, it could give them access to the consumer's systems and will result in unauthorized disclosure, data integrity and loss of data or service availability. [16]

*Recommended risk mitigation:*
SC-24 FAIL IN KNOWN STATE – Failure in known and secure state helps the organizations to prevent the loss of confidentiality, integrity, or availability of information when a system failure occurs. [17]

AR-4 PRIVACY MONITORING AND AUDITING – mitigates the treats of data loss or leakage by performing routine risk assessments and implementing appropriate fixes before a failure occurs. [17]
DM-2 DATA RETENTION AND DISPOSAL – implements a data retention policy as recommended by the National Archives and Record Administration (NARA). [17]

To prevent, detect and properly mitigate data loss or leakage in the cloud, AR-24, DM-2 and SC-24 can help consumers manage risks before they occur.

**Account or Service Hijacking (Risk level = 4B)**
Account or service hijacking is a threat that involves an attacker gaining unauthorized access to a cloud consumer's account. This vulnerability can cause the loss of confidentiality, integrity and availability of their data and services in the cloud; it usually occurs if the consumers do not properly manage authentication and authorization for their cloud services. Social engineering strategies like phishing, baiting, quid pro quo, pretesting can be used to obtain information about a user, which will enable the attacker to gain unauthorized access to the consumers account. With this access, the attackers can return falsified information, redirect consumers to illegitimate sites, temper with their data, and monitor or perform fraudulent financial transactions. [16]

*Recommended risk mitigation:*
AT-2 SECURITY AWARENESS – recommends regular security awareness for employees, so they can be aware of basic security risks and procedures. [17]
AC-2 ACCOUNT MANAGEMENT – identifies and setup the type of system accounts, roles, groups, functionalities required to support business operations. It also assigns account managers to oversee the authentication, authorization and termination of accounts in specific assigned systems. [17]
CM-7 LEAST FUNCTIONALITY – recommends configuring the least amount of functionalities, decoupling system components and disabling unneeded functions. [17]
CA-7 CONTINUOUS MONITORING – performs frequent continuous monitoring of systems and security status, regular security assessments, and performing corrective actions, in case any anomalies are detected. [17]

Performing the combination of AT-2, AC-2, CM-7, CA-7, and other controls, allow the consumers to properly implement a defense against account hijacking. In addition, the consumers should implement an email system that warns their user of external emails, to avoid phishing and other social engineering attacks.

**Unknown Risk Profile (Risk level = 3D)**
At the IaaS layer, it is important for the consumers to keep track of the various software versions, code updates, operating systems updates, security patches and intrusion attempts on their cloud services. However, consumers have no control and little to no visibility on how the cloud providers manage their internal security. [16] Often organizations are not aware of their risk profile, because of the following reasons: audit logs are not frequently monitored or reviewed, no baseline configuration, no proper configuration management to track changes, configuration and security setting are not periodically reviewed, quarterly CVEs (Common Vulnerability and Exploit) patches are not applied, and no routine risk assessments or penetration testing are performed.

*Recommended risk mitigation:*
CM-2 BASELINE CONFIGURATION – This control establishes and document baseline configurations for information systems and system components including connectivity-related, communication, integration points, security, systems status and more. [17]
CM-6 CONFIGURATION SETTINGS – This control identifies, documents, and monitors configuration settings like hardware, software, firmware and other systems components; it also approve changes in case an update is required. [17]
AU-12 AUDIT GENERATION – This control generates audit trial for different management and data driven events from across many information systems. These auto trial could include configuration changes, user logins, logs and others events that occurred at a particular time. [17]
CA-8 PENETRATION TESTING – This control helps organizations to perform specialized assessments to identify any possible vulnerability that could be exploited or determine to what extend their information system can resist any form of physical logical or cyber-attacks. [17]

Most large enterprise with enormous technology foot prints and traditional IT data centers, are already utilizing similar controls like the above recommendations, to avoid or mitigate unknown risk profiles. To properly mitigate these unknown risks profile in the cloud, consumers need to ensure that the providers are applying the appropriate controls like AU-2, CA-8, CM-2, CM-6 and security plans; when possible, request periodic uptime and system status reports from the SaaS and/or PaaS providers. Additionally, IaaS consumers have the responsibilities to frequently monitor and review audit logs, set baseline configurations, track change management, review security and configuration settings, apply CVE patches, and even request permission from the cloud providers to perform penetration testing in their cloud environments.

**Operational Risks (Risk level = 4D)**
Below are some operational gaps that could lead to the vulnerabilities and threats discussed in the proceeding section. The section will expand on these gaps and provide recommendation on how organizations need can properly plan before, during and after migrating their data and systems to public cloud environments.

*Implementing too quickly:*
Organizations can become more attracted by the capital cost savings, agility, speed to market and other benefits that cloud services promise, that they do not properly plan and evaluate before migrating their data and systems to the cloud. For example, those with higher need for resilience, availability and performance. [1]

*Integration issues:*
Without proper due diligence and project management standards, organizations may have a difficult time integrating their internal systems and data with cloud-based services. [1] For example, organizations with large amount of legacy technologies.

*Moving the wrong data or applications to the cloud:*
Consumers should be mindful when migrating critical applications or data to the cloud. Ideally, migrations should begin with non-critical programs and services. Only move mission critical data and application after the cloud environment has been fully designed and tested. [1] For example, customer facing data and applications.

*Liability:*
Organizations should be concerned about their business exposure in the event of a security breach, including any impact on their earnings, reputation, and customer relationships. These security breaches could also lead to the loss of integrity and availability of data and system components, as well as legal liabilities. [1] For example, government secrets or healthcare institutions with PII and ePHI regulations.

*Compliance:*
Organizations in highly regulated industries like banking, financial services, and healthcare may find that it is too risky to rely on the cloud. In such case, these consumers need to plan extensively and develop System Security Plan (SSP) and Risk Management Plan to mitigate all the cloud risks discussed in previous section. [1] For example, a federal government agencies high impact systems with need for FISMA and FedRamp compliance should consider using GovCloud services from Amazon or Azure.

*Cost implications:*
"Organizations that move their entire infrastructure to the cloud may find that they are paying more than if they owned all of the infrastructure themselves. At the same time, some organizations may find that their cloud solutions are costing them more than they expected because they did not properly budget for data replication, backup, and other support services". For example, moving entire infrastructure to the cloud may enterprise consumers millions of dollars a month in operational cost with may be a lot higher than a traditional IT data center.

### Historical analysis
Historically, cloud computing has proven to be more than on premise IT environments, because they go through frequent audits and must meeting different compliance standards to meet their consumers' needs.
Based on over 300,000 consumers surveyed, only about 9% had experienced cloud security incidents, while 55% had not, 15% could not disclose and 21% were unsure. [3]

After extensive research from most recent years, there are very few cloud computing data breaches, and majority were likely caused by the consumers handling of their shared responsibility portion of the agreement. Some examples were misconfiguration issues, mismanaged credentials or unprotected passwords, and others.

Below are summaries of cloud incidents that occurred in recent years.

### Cloud Data breach cause by providers:
*Microsoft Data Breach:*
In December 2010, Microsoft, the second largest cloud provider worldwide, experienced a data breach when some unauthorized users had access to Microsoft employees' information, when they downloaded the copies of the Business Productivity Online Suit. The issue was traced back traced back to an incorrect product configuration by the cloud provider (Microsoft), but it was realized and fixed within two hours. So there was no major damage. [12]

This incident could have been prevented if Microsoft had properly utilized proper development standards, processes and tools. They should have properly tested the software to ensure that there is no bug before releasing it to the public. See the recommendations under: "Insecure Interfaces and APIs (Risk level = 3C)" for details.

*Dropbox Data Breach:*
In mid-2012, Dropbox, the second largest cloud file sharing provider worldwide, experienced a data breach and the hackers gained unauthorized access to over 68 million customers information, which included emails and passwords, this information was also sold on the dark web in exchange for bitcoin. [12] The cause of this hack was traced back to an employee who re-used the same password for their LinkedIn account. When LinkedIn was hacked the hackers used that employee's password to login to their Dropbox admin portal.

The incident could have been prevented if Dropbox was conducting proper awareness training, separation of duties, and using firewall security to prevent outsiders from accessing the administrative aspect of their system, which should only be accessed from their internal network or through VPN tunnel. See the recommendations under: "Insider threat (Risk level = 3D)" for more details.

*Yahoo Data Breach:*
In 2013, Yahoo, one of the largest cloud email providers worldwide, experienced the largest data breach in history when over 1.5 billion users' email accounts were compromised in two separate data breaches. The disclosed data included Personal Identifiable Information (PII) like names, date of birth, email addresses, password, email contents and even sensitive security questions. The issue was traced back to a falsified cookie (session hijacking) that allowed the unauthorized users to gain access to Yahoo internal systems. [12]

The incident could have been prevented if Yahoo had properly implemented the combination of Concurrent Session Control, Session Lock, and Session Authentication to ensure the confidentiality and integrity of user's email accounts. See the recommendations in "Session Riding and Hijacking (Risk level = 4C)" for more details.

**Cloud Data breach caused by consumers:**
*Uber Data Breach:*
In October 2016, hackers gained access to Uber's, the largest ride share/ cloud transportation provider, AWS credentials and over 57 million customers and drivers' information was exposed to the hackers. This breach happened because Uber stored their AWS credentials in their GitHub repository, and the GitHub account did not have secure password protection. [11]

If Uber was conducting regular security awareness training, enforcing password policies and storing important passwords in protected password vaults, this incident would not have occurred. See the recommendations under: "Account or Service Hijacking (Risk level = 4B)" for additional details.

*Accenture Data Breach:*
In 2017, Accenture, one of the top 8 technical and corporate consulting firms worldwide, incorrectly configured 4 of the S3 storage to be accessible to the public. These buckets stored a lot of critical information, including: thousands of plain text passwords, and encryption keys. Serious data breach and damages would have resulted, if a malicious user had accessed those buckets from the public. [11]

If Accenture had baseline configuration, automated configuration settings, and proper audit trailing monitoring processes in place, mistakes like these would not have gone unnoticed. See the recommendation under "Unknown Risk Profile (Risk level = 3D)" for further details.

*Mexico Election Data Breach:*
In April 2016, the National Electoral Institute of Mexico experienced a data breach and over 93 million registered voters' information was compromised, that is over 70% of the entire country's population. The cause of the data breach was due to an insecure database configuration that exposed confidential information to the public. The database was hosted on the AWS cloud, in another country, which is non-compliant to host national records in another country. [12]

This could have been prevented, if the institution had an operational plan in place before migrating to the cloud. See "Operational Risks (Risk level = 4D)" for details.

## Recommendations for ensuring a secure Migration
### Not every data or application should be migrated to the cloud
Best practices suggest that organizations should not move Personally Identifiable Information (PII), Payment Card Industry (PCI), Electronic Protected Health Information (ePHI) or critical applications to the cloud, in the beginning stages of migration. Unless they have properly planned the migration and ready to trust the provider to keep their sensitive information secret in the beginning. [3]

### Properly plan and develop a full scope of work
Organizations need to invest time in developing a full scope of work for their migration, including requirements, technical resources, architecture, what will be migrated, timelines, and a security and risk management plan. The can be accomplished by expanding existing policies and procedures to the cloud environment or adopting a framework that works best for the new implementation. [3]

### Evaluate, Audit and Visit potential vendors
Once organizations have a scope of work, they should carefully vet potential vendors. They should consider issuing an RFP that asks questions about the organization's history, capabilities, and experience. When possible, visit on-site and audit the provider's current Standard Operating Procedures (SOPs) to confirm compliance and operational efficiency. [3]

### Start the migration with non-critical applications and data
Organizations that are new to the cloud need to start by migrating less critical applications or data. This will give them time to evaluate the migration process, and improve the process as necessary. Also ensure that a baseline configuration, a configuration management and change management processes are in place before migrating anything. [3]

### Pause and perform Quality Assurance (QA) during the migration process
Once the initial migration is complete, organizations should pause and validate everything before migrating more applications or data. During this stage, security, performance and availability should be the top goals in mind. [3]

### Keep everything secure during and after migration
Being organized and working with a well-defined project plan is a key part of a successful and secure cloud migration.
Organizations must be careful and keep both on premise and cloud environments secure during the migration. Applications may need different permission settings based on where they are residing, and firewalls may have to be configured differently, but proper planning can prevent a poorly configured cloud environment. [3]

**Recommendations for operational efficiency and Security**
Designing for operational efficiency depends on your organization's business needs and how the applications can adapt to the new environment. The rise of cloud computing has enabled organizations to be more agile, therefore the speed at which organizations are moving has increased the responsibilities and demand on development and operation teams.

To properly manage these demands while staying secure, it is recommended to the following best practices: Development Operations (DevOps) concept, Infrastructure as Code (IaC), Configuration Management (CM) tools and automated Continuous Integration (CI), Continuous Delivery (CD) processes, and other Information Security best practices like (Authentication, Authorization, Auditing, Patch management, and other Risk management frameworks) to ensure that the teams meet the high demands while keeping the organization's configurations in sync and remaining safe during the cloud environment operations. Well established organizations may be able to extend their existing IT governance and security policies to their cloud environment, but might be a need to make enhancements to make them relevant to the cloud environment, and small and medium organization may need to adopt or develop a cloud policy from scratch.

The below sub-topics will expand on the recommended concepts, and the proceeding section will provide a sample architecture to explain how these concepts can be applied.

*Develop Cloud Governance Framework*
Managing security in a cloud-based environment begins with the implementation of a comprehensive IT governance framework. "Organizations need to develop strict governance frameworks to ensure cloud infrastructure and operations are as secure - if not more secure - than traditional on-premise approaches to protect corporate data and critical systems." [2]

In addition to the consumers developing their own cloud-based governance framework and data management policies to fulfill their part of the shared security and operation responsibilities; the cloud providers are required to maintain a set of comprehensive governance frameworks, policies, procedures and other standard operating procedures to manage the actions of both their employees and consumers. Examples of the major cloud providers' policies include:

Amazon **Web** Services – AWS Service Terms [14], and AWS Cloud Security [8]
- These document describes how Amazon Web Services, the second largest cloud provider worldwide, protects their systems from various risks, and restricts how consumers are expected to use their services.

Microsoft Cloud Services – Azure Security [15], and Microsoft Services Agreement [9]
- Microsoft, the largest cloud (SaaS, PaaS, IaaS) provider worldwide, use these documents to protects their systems from various risks, and provides direction on how consumers are expected to use their services.

These governance framework must also address how the organization protect important data and system, maintain an update date inventory and compliance status of all computing resources, backup and recovery procedures, Disaster recovery strategies, risk management processes, incidence prevention, detection and response procedures, and others.

Examples of these cloud governance frameworks can be evaluated and adopted from several sources, including: the Cloud Security Alliance (CSA), The Trusted Computing Group, the Information Security

Audit and Control Association (ISACA), National Institute for Standards and Technology (NIST), Federal Risk Management Program (FedRAMP), Federal Information Security Management Act (FISMA) and more. [2]

*Manage Authentication*

For SaaS and PaaS authentication, only provide access to employees or project resources who need access to specific systems. All user account creation must go through an IT Service Management (ITSM) approval process before they are creates. For very large organizations with multiple on premise and SaaS applications, make sure to integrate these applications with existing Active Directory, where possible. This will be enable Single Signed-On (SSO) and it will be manageable for both the organization and employees.
- Users can easily manage their credentials without having to remember or write down too many SaaS & PaaS credentials.
- When a user is no longer on a project or with the organization, the IT administrators can easily disable their Active Directory account and email account, without having to email every manager or application administrator to disable the user's account. There is a big room for error if you try to disable individual users from each application.

For IaaS, ensure that users who need access to the cloud console or dashboard, should login with Multi-Factor authentication. For example, AWS console, Azure console, Google cloud console, IBM, Rackspace and other cloud dashboards.
Under no circumstances must the root account be used, unless absolutely required, otherwise, users should always login with individual accounts credential.
- Connecting to server instances and virtual machines in the cloud.
  - It is recommended to connect to these instances using Active Directory (AD) rather than connecting with a public (.pem) key or other default authentication method provide by the cloud providers.
  - Never connect to cloud instances directly from the internet. At the very least, keep all critical data and servers in a private subnet when possible, and use the below methods to connect to them.
    - Only allow Remote Desktop (RDP) and/or Secure Shell (SSH) access from an internal subnet. And create a few jump servers in the public subnets and allow RDP or SSH access from those jump servers, to the critical instances.
    - For a more secured environment, it is recommended to use intranet and VPN connection to access those cloud servers.
- See the sample architecture designs for details how this is handled.

*Limit Authorization*

Once the authentication process has been established for the cloud servers and other applications, the next stage is to manage how much users can do when they access these cloud systems.
It is recommended to give users the minimal access that enables them to perform their job functions, and use groups to manage privilege in the various systems.

- For SaaS and PaaS create custom application and/or middleware group, then associate that with the users AD accounts. For example, accounting, human resources, managers, sales, IT admin, IT support, development.
- For IaaS, in addition to advanced technical administrators and engineers, other less technical users might need server/instance level access to do their job, use AD group policy objects to determine what they can do when they are on the servers. For example, Testers, Developers, App Specialists and support teams.

*Audit and monitor before migration and during operations*
Most cloud providers already submit to third-party auditors and regulatory compliance agencies. Depending on the critical nature of the consumers' data and systems, they should review those audit reports and/or send their internal IT Auditors or security experts to verify the security if their service standards.

In addition to verifying that the provider is complying with all necessary regulations, consumers also need to perform regular audits of their shared responsibilities:
- For SaaS level, many applications will allow the consumers to enable auditing at the application level. At this point, it is recommended to enable auditing for both failed and successful logins, and regularly monitor the audit trials.
- For PaaS level, cloud consumers can enable both application and database level auditing on their applications, database and middleware layers.
- For IaaS level, there is a lot the cloud consumers can do to audit their infrastructure. Users must enable both successful and failed logins for all applications, databases, middleware, servers, as well as other services and the network level traffic.
Enabling the audit trail is just half the battle, so the cloud consumers must also enable advance monitoring to get alerts on abnormal activities and also configure automated intrusion detection and response to remediate threats.

*Patches management strategies*
Ensure that all patches are always up to date. Create a patch management process that checks for updates, missing patches, Common Vulnerability and Exploit (CVE) database, and alert IT Administrators when action is needed.
Test patches in a lower level environment then apply to higher level environments. For example, Dev, Test, and then Prod). A best practice is to designate a dedicated patch testing environment, but before even applying the patches to a functional environment.

*Implement Automation*
The use of automation technologies and processes helps cloud consumers to standardize their IT and cloud operations, it also greatly improves security, efficiency, reduces cost and many more benefits. Below are some approaches and concepts organizations to implement to accomplish automation in their cloud and other IT environments.

### DevOps practice
Traditionally, IT Operations (Ops) teams and software development (Dev) teams work independently, so provisioning, deploying and managing servers and other IT resources becomes bottleneck between these two team, in a high agility and fast pace environment.
Implementing the DevOps concept enables consumers to combine efforts and tools to properly collaborate between their Dev and Ops teams, during the software build, deploy, test and release processes.

### Utilize IaC for provisioning
The use of cloud infrastructure and other services as greatly reduce the time it takes to procure and provision new hardware, servers and other resources in a traditional IT environment. However, manually configuring cloud infrastructure services can also be time consuming, error prone and increase the risk of configuration drifts between various components in the cloud environment. For example, the baseline configuration of network interfaces, firewall, servers, middleware, databases and security levels.

By utilizing Infrastructure as Code (IaC), Consumers can script their cloud computing resources into machine readable codes; these codes can be launched to provision initial systems, and future systems. The advantage is that the setup will be faster and standardized across all system components within their cloud environments.
Remember to use a Vendor agnostic IaC tool, for more portability.

### CM for Infrastructure and systems management
Without configuration management processes and tools, it would be difficult, if not impossible, for the consumers' technical teams to properly manage the systems setup by IaC process to keep all systems configuration in sync and up to date. For example, patching all or specific systems, and maintaining compliance across all systems.
Implement a proper configuration management process and tool to ensure that the environment configuration are up-to-date, fully synchronized, and monitor for any configuration drift.

### Implement CI/CD code pipeline
In the midst of high agility and speed to market, producing high quality software in a cost efficient way is very important. Without the use of a well-designed and automated code deployment pipeline, it would be challenging to for organizations to meet the market demand while producing quality and secure products.

The implementation of an appropriate code deployment processes and tools can enable organizations to automatically monitor code repositories for error, provision containers, deploy code, run automated testing tools, verify security matrices and report on the status of a build on a scheduled or event driven basis, all without human intervention.

### *Routine Penetration testing*
Periodic penetration testing to scan for vulnerabilities and threats at every level of the cloud environment, is very important in identifying and mitigating risks. Then develop and implement appropriate counter measures for any possible threat that is discovered. The can also be implemented with the use of Configuration Management.

### Recommended Cloud Architecture design
Below is a sample architecture design that addresses mitigates the vulnerabilities and threats in cloud computing, then applies the various security standards and tools recommended in this research paper. This architecture can be adopted and modified for both Public and Hybrid cloud deployment models, and it is also best suited for the PaaS and IaaS service tiers.

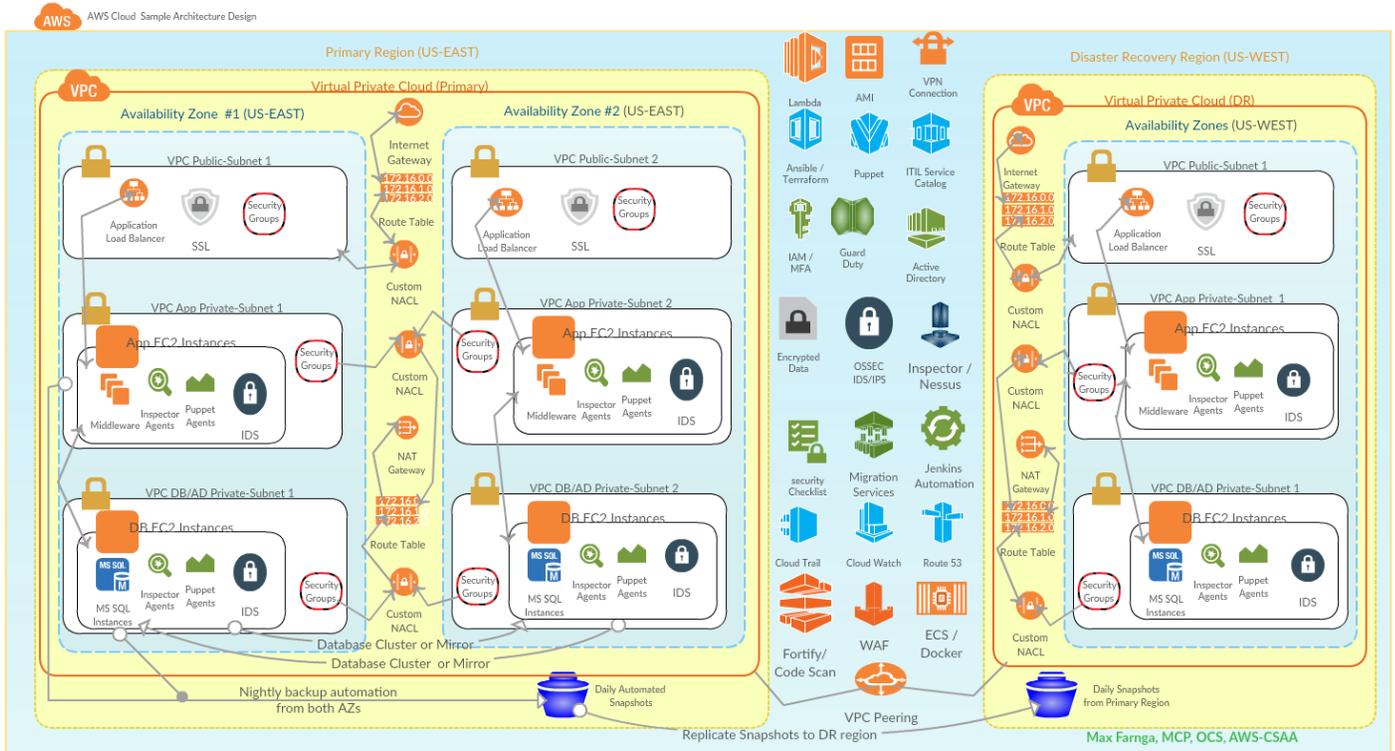

**Architecture overview:**
The above architecture is designed for high security, high availability, resilience, cost efficiency and disaster recovery. This architecture demonstrates a use case for a company that develops Software as a Service (SaaS) provider, and the hosts about 80 to 90 of their system on an AWS cloud infrastructure. Notice that the Risk management section uses FISMA controls and HIPAA standards to mitigate potential vulnerabilities and threats that could exist in a cloud environment.

Below is a description of the architecture and how it applies the various recommendations to protect organizations against the threats and vulnerabilities mentioned in this research.

*Virtual Private Cloud (VPC)*
VPC is a logically isolated network in AWS cloud, and the same exist with other cloud service providers. It can be compared to an individual data center network, except the consumer does not manage the physical aspect of the network.

The architecture includes a primary VPC in Virginia and a Disaster Recovery (DR) VPC in Oregon. Both VPCs should be designed using Infrastructure as Code (IaC), with the same baseline configurations and resources.

*Cost efficiency:*
Since it is very unlikely for the all 5 Amazon Data Centers to experience complete outage or for an entire state to be destroyed by a disaster, it is not cost efficient to pay for resources in the Disaster Recovery Region. So the DR VPC should be stored as a code, and that code should be updated whenever the primary VPC configuration settings are updated. And a daily backup should be shipped over to the DR region.

So in case of extreme Disaster, it will take a few minutes to rebuild the entire cloud architecture (network, active directory, firewalls, VPN and others) in the DR VPC, then restore the backup databases and applications, based on the daily backups.

### *Availability Zone (AZ)*
AZs are physically isolated locations (Data Centers) within a region.

The architecture recommends using multiple Availability Zones (at least 2 or more), and provision the Subnets and other cloud resources across these availability zones.
For example: The databases are clustered across at least two AZs and Elastic Load Balancers (ELBs) to balance the application loads across two or more AZs. In this case, when one AZ goes down, the applications and databases remain in operations with little or no interruptions.

### *Sub Network (Subnet)*
Subnets - are sub-sets of a larger Wireless Area Network (WAN) with logical sub-division of networks that can be geographically or categorically distributed.

The architecture uses subnets to determine which geographical location to provision infrastructure resources and also determines the level of security based on the critical nature of the resources. For example:
- Only Load Balancers are in the public Subnets with access to the internet.
- Applications, web, SFTP and other servers are in a private Subnets and can only be from the internet through the ELBs.
- Databases, Active Directory and other critical servers resides in a more secured subnet, and it can only be accessed from within the VPC.
- Both private subnets only access the internet through a NAT Gateway, only to download patches, and VPN tunnels to remote desktop (RDP) into the servers.

### *Network Access Control List (Network ACL or NACL)*
The Architecture uses NACL to manage custom firewall ports for network security to the various subnets. It only opens required ports and everything else is denied.

### *Security Groups (SG)*
The Architecture uses SGs to control network access to the various servers and other resources within the subnets. The second layer of the firewall and it only open required ports for resources.

### *Access and Identity Management (IAM)*
The Architecture uses IAM to manage authentication and authorization to various amazon cloud resources. Uses multi-factor authentication for users to access the cloud dashboard.

### *Active Directory (AD)*
The Architecture uses AD to manage authentication and authorization to the cloud servers and databases. Only internal uses have active directory accounts.

### *Amazon Machine Image (AMI)*
The Architecture uses AMI to manage standardized and approved Operating Systems images of baseline configurations. Only these baseline images are referenced in the IaC for provisioning.

### *GuardDuty*
The Architecture uses GuardDuty as a threat intelligence tool to manage Intrusions Detection and Prevention in the cloud. Consumers can also use other vendor's agnostic security tools like WebSAINT, ImmuniWeb, BeyondSaaS and Dell Secure Works. [13]

*Inspector*
The Architecture uses Inspector perform routine vulnerability scans and assessments. Nessus is a Vendor agnostic vulnerability assessment solution.

*Ansible*
The Architecture uses Ansible to manage Infrastructure as Code (IaC) for efficiency and cost effectiveness.

*Puppet*
The Architecture uses Puppet to manage automated configurations, patching, deployments and others.

*Cloud Trail*
The Architecture uses Cloud Trail to manage the auditing of management events and data driven events.

*Lambda*
The Architecture uses Lambda to run code on serverless infrastructure, which save cost because lambda on bills when the code actually runs, which is a lot better than having a server running 24/7 for infrequently accessed codes.

*Other security tools*
The architecture also uses other security tools like CloudWatch, Compliance Config and many other tools.

**Conclusion**

In recent years, cloud computing has become the most popular technology and it is increasingly becoming a household topic, because most of the world's population now have access to some sort of cloud services, through either SaaS, PaaS or IaaS service tier. Though cloud computing does present a lot of great benefits and advantages for individuals and businesses, many are unware of the risk landscape and how to protect their valuable data against the breach of Confidentiality, Integrity and Availability.

Despite the popularity of cloud computing worldwide; security has been the major barrier to cloud adoption among individuals and organizations with critical data. [6] For example, Government secrets, biomedical research, scientific and nuclear data, health records, financial information, personal identifiable information and classified data.

According to a 2016 survey of over 300,000 responds, 53% had security concerns, 42% had legal concerns, and 40% were concerned about data loss; as their barrier to cloud adoption. [3] This paper has helped consumers (individuals and organizations) to answer important questions like:
- How secure are the consumers' valuable data in the cloud?
- Should organizations migrate mission critical data and applications to the cloud?
- How can organizations design and implement cloud security framework and architecture?

Though cloud computing limits the consumers visibility and accessibility to the physical systems, a cloud environment can be very safe, or even secure than traditional IT environments, if the consumers have the right expertise and take their shared responsibilities very seriously.
Whether on premise or in the cloud, information systems are always a target for potential vulnerabilities and threats. In order to stay safe, organizations must develop a comprehensive cloud security framework, properly design a cloud architecture, and perform routine security assessments of their systems configuration, along with penetration testing, vulnerability scans, patch management and other recommendations provided in this research paper.